\documentclass[%
 reprint,
nofootinbib,
 amsmath,amssymb,
 aps,
]{revtex4-2}

\usepackage{graphicx}
\usepackage{dcolumn}
\usepackage{bm}

\begin{document}

\preprint{APS/123-QED}

\title{How many pixels are there in a polarized pulsar timing array map?}

\author{Dylan L. Jow}
\homepage{D.L.J. (author one) contributed equally to this work with A.T. (author two)}
\email{dylanjow@stanford.edu}
 \affiliation{Kavli Institute for Particle Astrophysics and Cosmology, Stanford University, 452 Lomita Mall, Stanford, CA 94305-4085, USA.}
 
\author{Anna Tsai}%
 \email{anna.tsai@mail.utoronto.ca}
\affiliation{Canadian Institute for Theoretical Astrophysics, University of Toronto, 60 St. George Street, Toronto, ON M5S 3H8, Canada}
\affiliation{Department of Physics, University of Toronto, 60 St. George Street, Toronto, ON M5S 1A7, Canada}%

\author{Ue-Li Pen}
\affiliation{
 Canadian Institute for Theoretical Astrophysics, University of Toronto, 60 St. George Street, Toronto, ON M5S 3H8, Canada}
\affiliation{Department of Physics, University of Toronto, 60 St. George Street, Toronto, ON M5S 1A7, Canada}
\affiliation{Dunlap Institute for Astronomy \& Astrophysics, University of Toronto, AB 120-50 St. George Street, Toronto, ON M5S 3H4, Canada}
\affiliation{Institute of Astronomy and Astrophysics, Academia Sinica, Astronomy-Mathematics Building, No. 1, Section 4,
Roosevelt Road, Taipei 10617, Taiwan}
\affiliation{Perimeter Institute for Theoretical Physics, 31 Caroline St. North, Waterloo, ON, Canada N2L 2Y5}
\affiliation{Canadian Institute for Advanced Research, CIFAR program in Gravitation and Cosmology, MaRS Centre, West Tower
661 University Ave., Suite 505
Toronto, ON M5G 1M1 Canada}

\date{\today}

\begin{abstract}
The standard Hellings and Downs approach to searching for gravitational wave signatures in pulsar timing array (PTA) data has successfully produced evidence for the presence of nanohertz-wavelength gravitational waves. However, it does not, on its own, produce any directional information and is insensitive to polarization. An alternative approach is to construct maps of the polarized power on the sky. In this paper, we present a simple quadratic estimator of the gravitational wave power as a function of direction on the sky that is sensitive to the polarization state of the wave. We describe the full, $S_2 \times S_2$, state-space of a polarized gravitational wave background across the sky and the Poincaré sphere describing polarization. A natural question arises from this perspective: what is the resolution of a polarized sky-map, i.e. effectively how many independent pixels can a such a map contain? In other words, how many distinct gravitational waves can a PTA distinguish? It turns out the answer is finite, and is approximately $N_{\rm res} = 16 \times 2 = 32$ per frequency, where 16 is the number of  resolvable sky-positions and 2 is the number of distinct polarization states, corresponding to an angular resolution of $58^\circ$, achieved with a PTA with more than $N_{\rm pulsar} \gtrsim 20$ pulsars.  We demonstrate that the variance of the map is equivalent to the HD significance, while for a single point source, a 3-$\sigma$ HD signal corresponds to a 5.2-$\sigma$ map significance.
\end{abstract}

\maketitle


\section{\label{sec:intro}Introduction} 

With recent evidence for a nanohertz gravitational wave background (GWB) \citep{NANOGravdetection, IPTAdetection, EPTAdetection, CPTAdetection, Parkesdetection}, we have entered into the era of ultra-low frequency gravitational wave astronomy. Evidence for the nanohertz GWB comes primarily from the observation of the expected Hellings-Downs correlations between the timing residuals of pairs of pulsars in a pulsar timing array (PTA). This angular, two-point correlation, however, is insensitive to many properties of interest that the GWB may possess, such as anisotropy and polarization. While it is not possible to measure the overall chirality of a truly stochastic and statistically isotropic background with a PTA \citep{Kato2016, Belgacem}, anisotropies in the polarization may be of interest \citep{Kumar2024, SatoPolitoKamionkowski2024}. Moreover, the astrophysical component of the background is expected to comprise an ensemble of nearly monochromatic and fully polarized sources arising from super massive black hole binaries. Thus, moving beyond the paradigm of background detection towards single-source detection will require an understanding of polarization-sensitive observables. Likewise, anisotropy in the background will inform our understanding of the physical origins of the signal, possibly distinguishing between astrophysical and cosmological origins \citep{SatoPolito2024}. Anisotropy, therefore, is a natural next target for PTA experiments. 

Map-making techniques, i.e. quadratic estimates of the signal amplitude as a function of angular direction on the sky, are a natural way of measuring anisotropies in the background and searching for bright, individual point sources. Map-making techniques have long been a standard and powerful part of the CMB community's toolkit, enabling many high-level analyses. While maps have been utilized in recent PTA analyses \citep{Nanograv2023Anisotropy, Lemke2024, Grunthal2025}, the technique is underdeveloped in the PTA space. In particular, further study of the resolution and noise properties of PTA maps is needed in order to use map-based techniques for robust detections of anisotropies and individual sources. In addition, map-based analyses of PTA observations have thus far neglected polarization: a key source of astrophysical information.

In this paper, we introduce a simple extension to PTA map-making procedures: simple quadratic estimators of not only the total intensity of the signal over the sky, but also the polarization state. Instead of creating a single, polarization \textit{insensitive} map by computing the polarization-summed response of the PTA, we compute two complex maps generated from the distinct PTA response to left- and right-circularly polarized sources. The maps can then be combined to generate full Stokes $I, Q, U,$ and $V$ maps over the sky. We can go further and, for each spatial pixel, generate a second map corresponding to an estimate of the power in the GWB in that direction, distributed over the Poincar\'e sphere, describing the source's polarization state. In other words, we can estimate the power in the GWB over the full $S_2 \times S_2$ space that describes both source position on the sky and polarization state. 

A natural question that arises from this framework is: what is the effective resolution, across both sky and polarization space, of a PTA? This question is highly relevant for map-making techniques because it is equivalent to asking how many effectively independent pixels a polarized map of the gravitational wave sky has. Cast yet another way, the central question of this paper is: how many distinct, polarized point sources can a PTA distinguish? Naively, one might assume that an infinitely dense PTA would have a resolution solely limited by the diffraction limit. This, however, is only true when the distances to the pulsars in the PTA are known, so that both the Earth term and pulsar term of the timing residuals can be measured. In the absence of well-constrained distances, it turns out there is a fundamental resolution limit that is independent of wavelength and is, in general, much larger than the diffraction limit\citep{BoylePen2012,2017ApJ...835...21R}.  Following a simple calculation, we find that for a dense PTA with poorly constrained distances to its pulsars, the number of distinct sources, per frequency, than can be resolved is finite and independent of frequency: $N_{\rm res} = 16 \times 2$, with $16$ distinct positions on the sky and $2$ distinct polarization states. In the language of map making, even an infinitely dense PTA effectively has only $16$ independent spatial pixels, each with $2$ independent polarization pixels. This corresponds to a maximum angular resolution of $58^\circ$, or $\ell_{\rm max} \approx 12$. Moreover, we find that this asymptotic resolution is achieved with a relatively modest number of pulsars, $N_{\rm pulsar} \gtrsim 20$. Note, however, that this fundamental limit on the resolution is only for a PTA for which the distances to the pulsars are unknown (so that the pulsar term is neglected in the analysis). Thus, searches for anisotropy at $\ell > 12$ will necessitate well-constrained distances. Without distances, adding pulsars to the PTA will only impact the signal-to-noise, but cannot improve the resolving power of the array.

The paper proceeds as follows: in Section~\ref{sec:observables}, we briefly describe the response function of a PTA to a polarized, monochromatic source of gravitational radiation. In Section~\ref{sec:maps}, we describe how this linear response can be inverted to produce maps of the gravitational wave power on the sky, as well as a spatial, pixel-by-pixel estimate of the polarization state. In Section~\ref{sec:resolution}, we perform simple simulations to compute the point-spread function (PSF) of a monochromatic source and the two-point correlation of noise maps. Using this, we arrive at the main result of the paper: even an infinitely dense PTA has a finite spatial and polarization resolution, when only the earth-term is taken into account. Section~\ref{sec:HD} compares the information content of a map with the standard Hellings-Downs analysis, and Section~\ref{sec:discussion} further discusses the importance of these results in the current observational landscape. 

\section{\label{sec:observables}The pulsar timing array response}

Here we give a brief overview of the observed response of a PTA to a gravitational wave source. For this paper, we will restrict out attention to purely monochromatic sources. The metric perturbation of such a source propagating from the $\hat{n}$-direction, with some frequency $f_0$, is given by
\begin{align}
    h_{ij}(t, x) = h \epsilon_{ij} e^{-2 \pi i f_0 (\hat{n} \cdot {\bf x} + t - t_0)}.
\end{align}
Note that here $\hat{n}$ points in the direction of the source, not in the direction of propagation. The polarization tensor, $\epsilon_{ij}$, is a normalized linear combination of the $+$ and $\times$ modes:
\begin{align}
    \epsilon^+ = \hat{\phi}_i \hat{\phi}_j - \hat{\theta}_i \hat{\theta}_j, \\
    \epsilon^\times = \hat{\theta}_i \hat{\phi}_j + \hat{\phi}_i \hat{\theta}_j,
\end{align}
where $\hat{\theta}$ and $\hat{\phi}$ are the spherical-coordinate basis vectors at $\hat{n}$. Alternatively, we can describe the polarization states in terms of the left- and right-circular polarized modes: $\epsilon^L_{ij} = (\epsilon^+_{ij} + i \epsilon^\times_{ij}) / \sqrt{2}$, $ \epsilon^R_{ij} = (\epsilon^+_{ij} - \epsilon^\times_{ij}) / \sqrt{2}$. This is the more natural basis for gravitational wave polarization, as astrophysical sources of gravitational radiation will primarily be circularly polarized. For a ring of test particles perpendicular to the direction of propagation, the left-handed polarization mode rotates the particles in a clockwise sense, as viewed from the origin. 

The effect of the metric perturbation is to produce a relative delay in the arrival time of the pulse from a pulsar located at a position, ${\bf r}$, known as the timing residual. The timing residual is given by \citep{Maiorano2021}
\begin{align}
    \tau(t; {\bf r}) &= \int df \tilde{\tau}({\bf r}) e^{-2 \pi i f t}, \\
    \tilde{\tau}(f; {\bf r}) &= \frac{i \tilde{h}_{ij}(f) \hat{r}^i \hat{r}^j}{2 \pi f (1 + \hat{n} \cdot \hat{r})} \left( 1 - \mathcal{P}(f; \bf{r})\right),
\end{align}
where $\tilde{h}_{ij}(f)$ is the Fourier transform of the wave, and $\mathcal{P}(f; {\bf r}) = e^{2 \pi i f r (1 + \hat{n} \cdot \hat{r})}$ arises from the so-called ``pulsar term" of the timing residual. Essentially, the total timing residual is the sum of the effect of the wave at Earth and at the pulsar. The pulsar sees the wave at a relative phase of $2 \pi i f r (1 + \hat{n} \cdot \hat{r})$ compared to the earth. 

We will be interested in monochromatic waves for which $\tilde{h}(f)$ is a delta function. We will also assume that the distances in the PTA are poorly determined, so that we neglect the oscillatory pulsar term, $\mathcal{P}$. For PTAs with unknown pulsar distances, the pulsar term may be treated as a source of noise in the limit where $2 \pi f r$ is large. In this limit, $\mathcal{P}$ is highly oscillatory and any two-point correlations of the timing residuals due to the pulsar term effectively vanish. 

Now, the Earth term only depends on the angular position $\hat{r}$ of the pulsar on the sky. Thus, we can compute the PTA response of a pulsar at position $\hat{r}$ on the sky for a circularly polarized wave propagating in the $\hat{n}$ direction:
\begin{align}
    &\tilde{\tau}^{R,L}(\hat{r}) = -\frac{h}{4 \pi f_0} \zeta^{R,L}(\hat{r};\hat{n},t_0), \\
    \begin{split}
    &\zeta^{R,L}(\hat{r};\hat{n},t_0) = e^{2 \pi i f_0 t_0} \,
    \times \\
    &\frac{\left(\cos\theta \sin \theta_{\rm gw} - \cos \theta_{\rm gw} \cos \left( \phi - \phi_{\rm gw} \right) \sin \theta \pm i \sin \theta \sin \left( \phi - \phi_{\rm gw}\right) \right)^2}{\cos\theta \cos \theta_{\rm gw} + \cos\left(\phi - \phi_{\rm gw} \right) \sin\theta \sin \theta_{\rm gw} - 1},
    \end{split}
\end{align}
where $(\theta, \phi)$ are the angular coordinates of the pulsar on the sky, and $(\theta_{\rm gw}, \phi_{\rm gw})$ are the angular coordinates of the gravitational wave source. We have also defined the dimensionless angular response, $\zeta^{R,L}(\hat{r};\hat{n},t_0)$, which is independent of the amplitude of the wave.

\begin{figure}
    \centering
    \includegraphics[width=\linewidth]{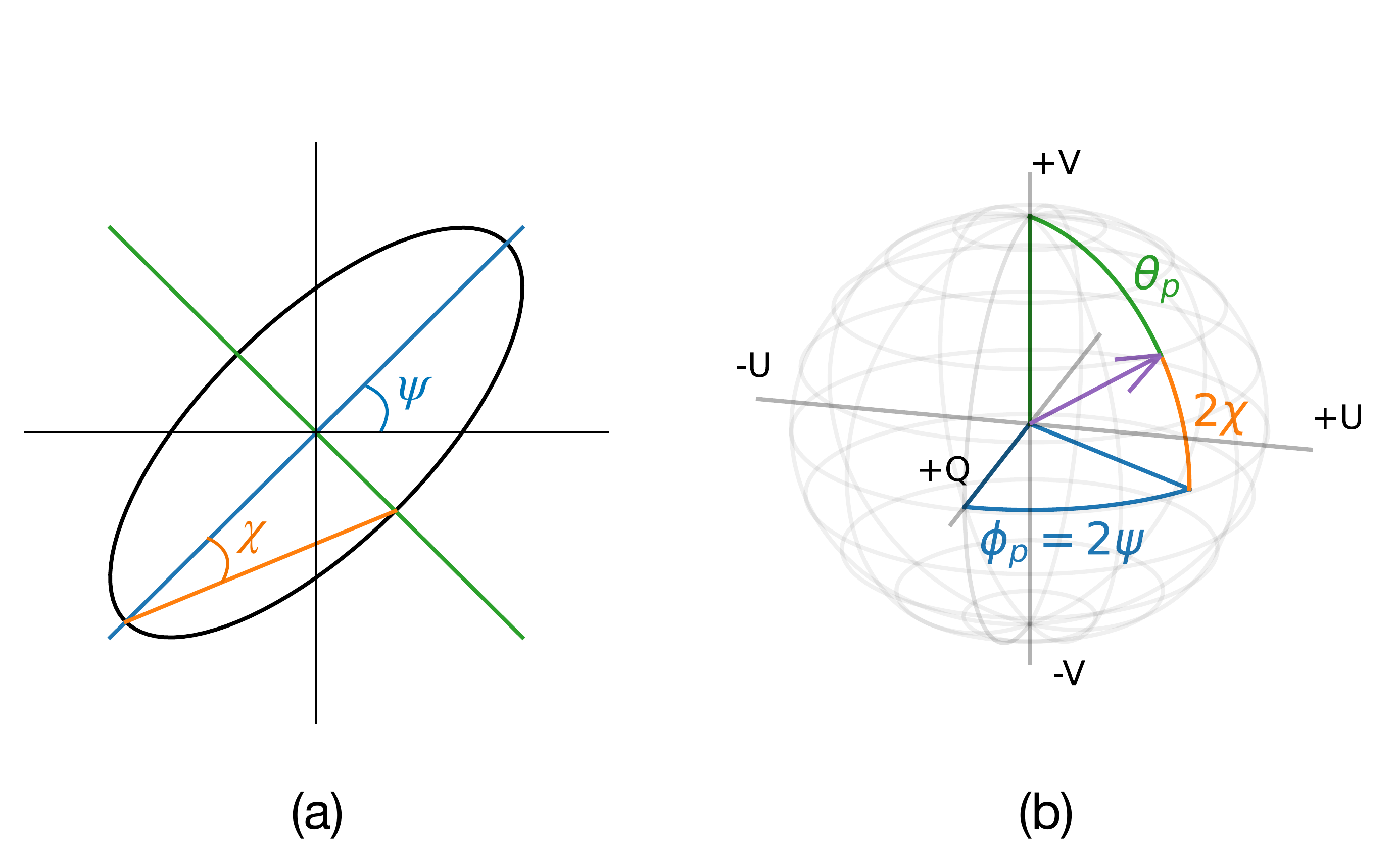}
    \caption{Diagram of the polarization angles that specify the polarization state. (a) shows the relation between the angles $(\psi, \chi)$ and the polarization ellipse, and (b) shows the angles on the Poincar\'e sphere and their relationship to $(\psi, \chi)$.}
    \label{fig:Stokes_diagram}
\end{figure}

Now, let us consider a gravitational wave source with an arbitrary polarization. The polarization state is described by the angles, $(2\chi, 2\psi)$, which define the polarization ellipse shown in the diagram, Figure~\ref{fig:Stokes_diagram}. The angle $\psi$ defines the direction of the linear polarization, and $\chi$ describes the degree of circular polarization. An alternative description of polarization is the Poincar\'e sphere, also shown in Figure~\ref{fig:Stokes_diagram}. A point on the Poincar\'e sphere, defined by the unit vector $\hat{n}_{\rm p} = (\sin\theta_p \cos\phi_p, \sin\theta_p \sin\phi_p, \cos\theta_p )$, is related to the polarization angles by $\theta_p = \frac{\pi}{2} - 2 \chi$ and $\phi_p = 2\psi$. The axes of the Poincar\'e sphere correspond to the polarization states described by the Stokes parameters: $Q$, $U$, and $V$. Using the polarization angles, we can write the general PTA response for a polarized, monochromatic gravitational wave as:
\begin{align}
\begin{split}
    \zeta(\hat{r};\hat{n},\hat{n}_p, t_0) = &\sin(\theta_p/2) e^{-i \phi_p / 2} \cdot \zeta^L(\hat{r}; \hat{n}, t_0) \\
    &+ \cos(\theta_p / 2) e^{i \phi_p / 2} \cdot \zeta^R(\hat{r}; \hat{n}, t_0).
\end{split}
\end{align}
The timing residuals are obtained by $\tilde{\tau} = \zeta h / 2 \pi f_0$. From this, we can see that the gravitational wave source is fully described by its position $\hat{n}$, its polarization state $\hat{n}_p$, and its initial phase $2 \pi f_0 t_0$. For the purposes of map making, the phase will turn out to be irrelevant, and so, the full state-space is $S_2 \times S_2$. Our goal is to construct a quadratic estimate of the power at each point across this state space.

\section{\label{sec:maps}Mapping the sky with PTAs}


Making a map of the gravitational wave sky boils down to answering the question: given an observed PTA response, what is the distribution of sources on the sky, along with the polarization state of the sources, that could produce such a response? Many methods can be employed to answer this inverse problem, but quadratic estimators remain the simplest. 

Consider a PTA with $N$ pulsars, located at $\hat{r}_i$, for $(i =1, ..., N)$, on the sky. Let $\hat{n}_j$ be a set of $M$ source-positions, uniformly distributed on the sky. We construct two $N \times M$ aperture matrices:
\begin{align}
    A^L_{ij} &= \zeta^L(\hat{r}_i;\hat{n}_j,\hat{n}_p=+\hat{z}), \\
    A^R_{ij} &= \zeta^R(\hat{r}_i;\hat{n}_j,\hat{n}_p=-\hat{z}).
\end{align}
The columns of $A^{L(R)}$ are the response of the PTA for a left (right) circularly polarized wave at a position $\hat{n}_j$ on the sky. The normalization is chosen so that, in the resulting intensity map that we will define later, the pixel at $\hat{n}_j$ has $I_j = h / 2 \pi f_0$ for a single gravitational wave source located at $\hat{n}_j $.  

Now, let ${\bf d} = [\tilde{\tau}_1,...,\tilde{\tau}_N]$ be the observed response of the PTA, where $\tilde{\tau}_i$ is the Fourier transform of the timing residual for the $i^{\rm th}$ pulsar. Also, let $\Sigma$ be the $N\times N$ covariance matrix characterizing the noise of the pulsar timing residuals. We construct the left and right ``dirty maps":
\begin{align}
    {\bf L} &= (A^L)^\dagger \Sigma^{-1}{\bf d},\\
    {\bf R} &= (A^R)^\dagger \Sigma^{-1} {\bf d}.
\end{align}
When $\Sigma$ is diagonal, the $M$-dimensional vectors, ${\bf L}$ and ${\bf R}$, simply describes the noise-weighted overlap between the observed response and the source distribution on the sky for a purely-left and purely-right background, respectively. In other words, the element $L_j$ is the inner-product between the observed PTA response, ${\bf d}$, and the response that would have been observed had the source been a left-circularly polarized wave originating from $\hat{n}_j$. Thus, ${\bf L}$ and ${\bf R}$ are quadratic estimates of the amount of left and right power in the sky as a function of direction. One might alternatively describe the aperture matrices $A^{L, R}$ as matched filters for the pulsar response generated by polarized point sources, and ${\bf L}, {\bf R}$ as the matched-filter response.

Now, ${\bf L}$ and ${\bf R}$ are complex, but can be converted to the more familiar Stokes parameters via
\begin{align}
    I_j &= |R_j|^2 + |L_j|^2,\\
    Q_j &= 2 {\rm Re}\left[L_j^* R_j\right], \\
    U_j &= 2 {\rm Im}\left[L_j^* R_j\right], \\
    V_j &= |R_j|^2 - |L_j|^2.
\end{align}
The result is four, real maps that describe the power distribution on the sky in the different polarization states. The intensity map, $I_j$, estimates the total power (regardless of polarization) coming from $\hat{n}_j$. Note that these four maps are not independent, as $I^2 = Q^2 + U^2 + V^2$ for a fully polarized source\footnote{Our restriction to monochromatic waves implies that the sources are fully polarized, so that this relation holds. Moreover, astrophysical sources of nanohertz GWs are expected to be fully polarized.}. Note also that none of the Stokes parameters depend on the complex phase of the observed PTA response, ${\bf d}$, nor would they have depended on the phase $2 \pi f_0 t_0$, had we not initially set $t_0 = 0$. This is unsurprising, since the Stokes parameters describe the polarization state of the wave regardless of its phase. However, readers wondering how we went from two-complex maps ${\bf L}$ and ${\bf R}$ (which are four independent, real maps) to three independent, real maps should note that the information we have lost is precisely the phase information. 

In standard map-making analyses, the ``dirty maps" are a step towards the construction of ``clean maps". The clean maps are the maximum likelihood estimates of the power distribution on the sky, maximizing a chi-squared likelihood function given the observed response \citep{2001A&A...374..358D, Dodelson, Grunthal2025}. To construct polarized clean maps, we must introduce a third, $N \times 2M$, aperture matrix, which is a concatenation of the left and right aperture matrices defined above:
\begin{align}
    A &= [ A^L, A^R].
\end{align}
The left and right dirty maps are simply the first $M$ and last $M$ elements of the vectors $A^\dagger \Sigma^{-1} \mathbf{d}$. The \textit{clean} left and right maps, $\mathbf{\mathcal{L}}$ and $\mathbf{\mathcal{R}}$, are given by
\begin{equation}
    \begin{bmatrix}
        \mathbf{\mathcal{L}}\\
        \mathbf{\mathcal{R}}
    \end{bmatrix} =
    (A^\dagger \Sigma^{-1} A)^{-1} A^\dagger \Sigma^{-1} \mathbf{d} 
    = (A^\dagger \Sigma^{-1} A)^{-1}
    \begin{bmatrix}
        \mathbf{L}\\
        \mathbf{R}
    \end{bmatrix}.
\end{equation}
We can then construct clean versions of the $I$, $Q$, $U$, and $V$ maps in the same way as above. The effect of inverting by the covariance matrix $(A^\dagger \Sigma^{-1} A)^{-1}$ is to de-convolve the dirty map with instrument response, i.e. the PTA point-spread function (in both angular and polarization space). This can be easily seen when one considers an observed response from a purely circularly polarized point source in the direction of one of the pixels, $\hat{n}_j$. That is, assume the actual observed response is given by $\mathbf{d} = A \mathbf{s}$, where $s_j = 1$ for a left wave (or $s_{2M + j} = 1$ for a right wave), and zero otherwise. Then, if there is no measurement noise, the true sky and polarization distribution is recovered exactly: $[\mathcal{L}, \mathcal{R}] = \mathbf{s}$.

Since the primary focus of this paper is to understand the limiting resolution properties of PTA maps, we are primarily interested in computing the dirty maps, instead of the clean maps. It is precisely the matched-filter response of the dirty maps to a point source (that one tries to de-convolve to produce the clean maps) that determines the resolution of a PTA in the presence of noise. 

\subsection{\label{sec:pcsphere}The Poincar\'e sphere} 

We have, thus far, outlined a procedure for constructing three, independent maps which estimate the power in Stokes $Q$, $U$, and $V$ on the sky. However, as we have noted, the full state-space of the gravitational wave source is $S_2 \times S_2$. 

Let $\hat{n}^{p}_{k}$ be a set of $K$ unit-vectors distributed uniformly on the Poincar\'e sphere. We construct a third estimator from the polarization aperture matrix:
\begin{align}
    B_{ik}(\hat{n}) = \zeta(\hat{r_i};\hat{n},\hat{n}^{p}_{k}).
\end{align}
From this, we construct another dirty map,
\begin{align}
    {\bf S}(\hat{n}_j) = B^\dagger(\hat{n}_j) \Sigma^{-1} {\bf d},
\end{align}
where $|S_k|^2$ estimates the distribution of power over the Poincar\'e sphere, at the spatial pixel $\hat{n}_j$. Thus, we have a map, $I_j$, which described the total estimated power at each position, $\hat{n}_j$, on the sky, and, for each $\hat{n}_j$, we have a map, $|S_k|^2$, which estimates how much of that power is in the polarization state described by $\hat{n}^{p}_{k}$. We can consider $\left\{ \hat{n}_j | \, j=1,...,M \right\}$ and $\left\{ \hat{n}^{p}_{k} | \, k=1,...,K \right\}$ to be a full pixelation of the $S_2 \times S_2$ state-space. Thus, our maps comprise a total of $M \times K$ pixels. In constructing these maps, we had the freedom of choosing $M$ and $K$. The central question of this paper is whether there is a physical limit to how many independent pixels such maps could, in principle, contain. This is equivalent to asking how many distinct gravitational wave sources a PTA could possibly distinguish.

\section{The PSF and map resolution}
\label{sec:resolution}

\begin{figure}
    \centering
    \includegraphics[width=\linewidth]{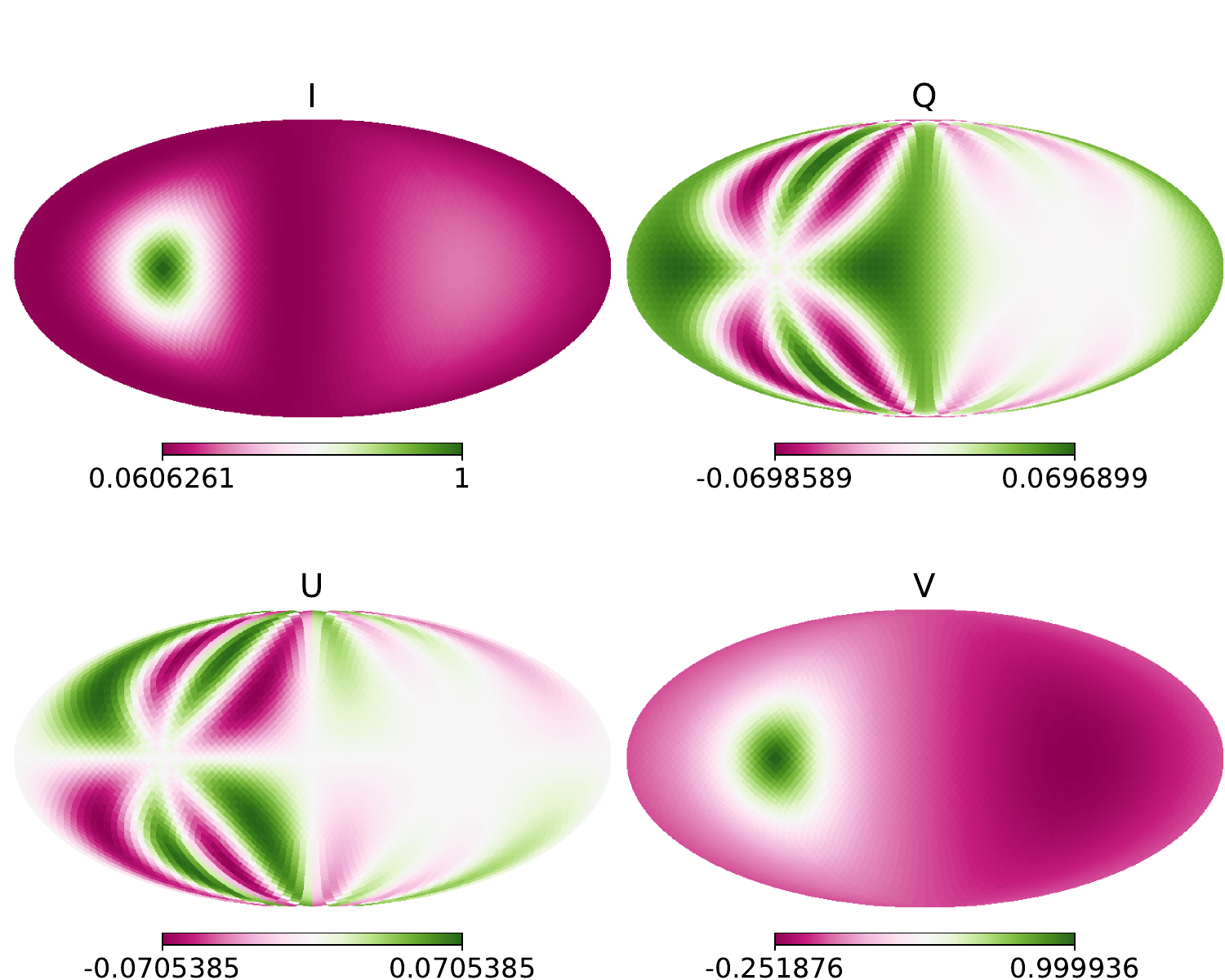}
    \caption{The matched-filter response of a purely right circularly polarized gravitational wave point-source located at $\hat{n} = \hat{y}$. The $I, Q, U$ and $V$ maps are constructed for a dense PTA with $N = 1200$ pulsars distributed uniformly on the sky. The maps use the \texttt{HEALPix} pixelization scheme, with $N_{\rm side} = 20$, or $M = 4800$ pixels. }
    \label{fig:IQUV_psf}
\end{figure}

In order to assess the resolution of a PTA, we need to know the response of the maps to a polarized point-source. Figure~\ref{fig:IQUV_psf} shows the resulting dirty map of a purely circularly-polarized point-source emanating from $\hat{n} = \hat{y}$ and produced by a dense PTA with $N = 1200$ pulsars, distributed roughly uniformly on the sky. The nominal number of pixels in our map, $M$, can be freely chosen independent of the PTA configuration. We use the \texttt{HEALPix} \citep{HEALPix} pixelization scheme to distribute the pixels roughly uniformly on the sky, and, for Figure~\ref{fig:IQUV_psf}, we choose $N_{\rm side} = 20$, which results in $M = 4800$. We have also assumed that $\Sigma \propto \mathbb{I}$, the identity matrix, so that the effect of noise to simply multiply the pixels uniformly by a constant. The resulting dirty map is the matched-filter response for a right circularly-polarized gravitational wave point-source in Stokes $I$, $Q$, $U$, and $V$. For a circularly-polarized source, the resulting maps only have significant power in $I$ and $V$. Note that we have also normalized the maps so that the maximum pixel in intensity has a value of unity.  From this, we can immediately see that even for a dense PTA, there is a lower limit to the achievable resolution: the power in $I$ remains distributed over a large angular region about the maximum. We will characterize the size of the PTA response by computing the solid angle on the sky such that $I > 0.5\,I_{\rm max}$, yielding $\Omega^{I}_{\rm p.s.} = 0.55\,{\rm sr}$, corresponding to an angular radius of $24.2^\circ$. Note, however, this should not be interpreted as the angular resolution of the PTA without first accounting for the noise properties of the dirty maps, as we discuss below.

\begin{figure}
    \centering
    \includegraphics[width=\linewidth]{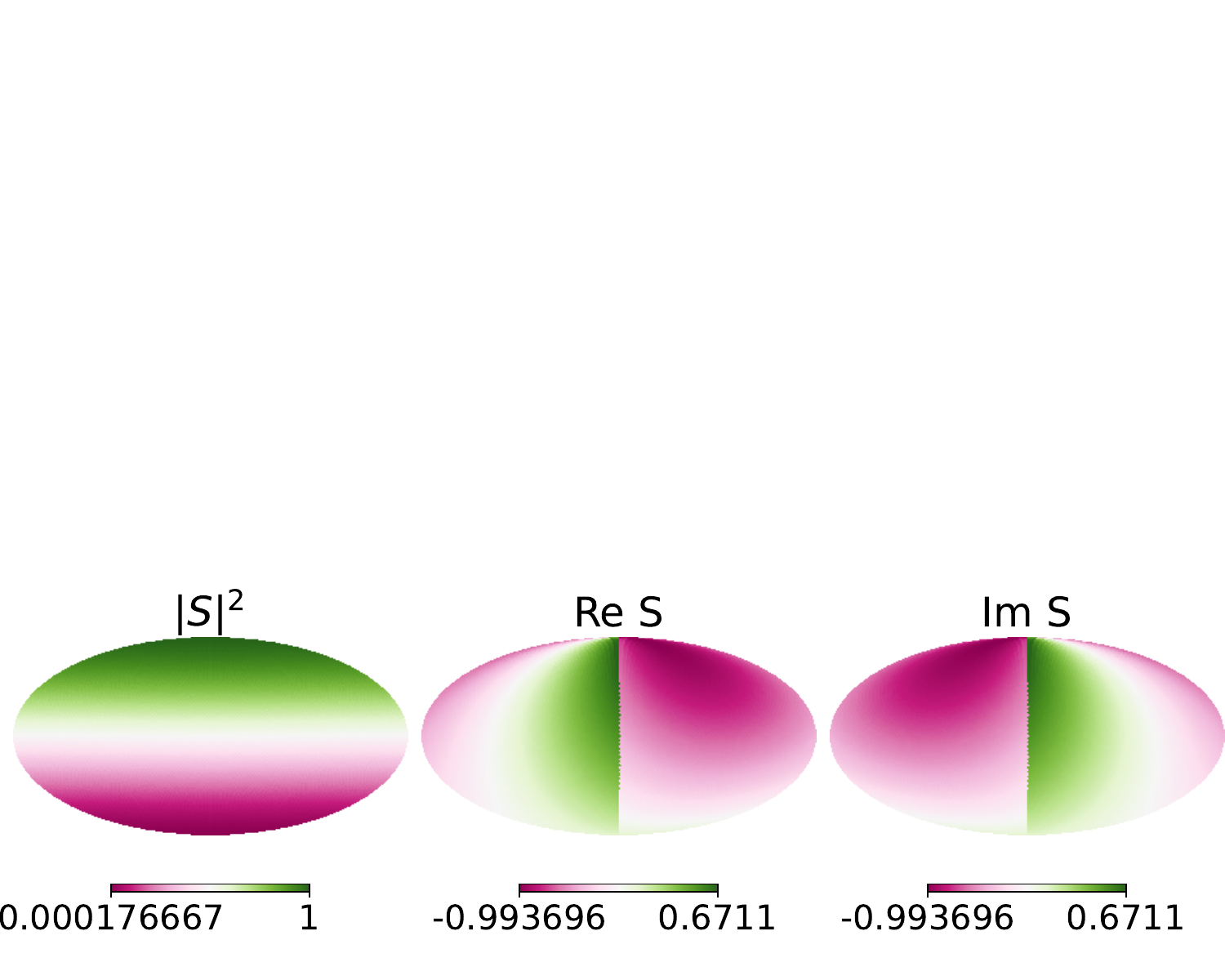}
    \caption{The Poincar\'e sphere dirty map for a purely right circularly polarized point-source located at $\hat{n} = \hat{y}$. The Poincar\'e map is produced for the spatial pixel corresponding to $\hat{n}_j = \hat{y}$ (the true gravitational wave direction).}
    \label{fig:pc_psf}
\end{figure}

Since our map making procedure is sensitive to polarization, we would also like to characterize the size of the map response in polarization space.  As described in Section~\ref{sec:pcsphere}, for each sky-pixel, $\hat{n}_j$, we construct a second map of the polarization state, described by the Poincar\'e sphere. Figure~\ref{fig:pc_psf} shows the reconstructed Poincar\'e sphere for the same source and PTA configuration as Figure~\ref{fig:IQUV_psf}, conditioned on $\hat{n}_j = \hat{y}$ (the true direction of the gravitational wave source)\footnote{As can be seen in Figure~\ref{fig:pc_psf}, there is a discontinuity along $\phi_{\rm pc} = 0$. This discontinuity arises from the fact that the polarization vector, defined by the angle $\psi = \phi_p / 2$, undergoes only one $\pi$-rotation for every $2 \pi$-rotation in $\phi_p$. Thus, the polarization vector undergoes a change in sign due to a full $2 \pi$-rotation in $\phi_p$, leading to the discontinuity. However, since $\pi$-rotations of the polarization vector represent the same polarization state (in fact, for gravitational radiation, $\pi/2$-rotations represent the same polarization state), this change in sign only represents a change in the phase of the wave, and, thus, the $|S_k|^2$ map does not suffer from any discontinuity.}. Again, we are free to choose the pixelization of the map, and we choose a \texttt{HEALPix} $N_{\rm side} = 20$, leading to a total $K = 4800$ polarization pixels. Computing the fraction of the sky for which $|S_k|^2$ exceeds half its maximum yields $\Omega^{\rm pol}_{\rm p.s.} = 2 \pi$, or half the sphere. In principle, we must compute the size of the polarized point-source response over the Poincar\'e sphere for every spatial pixel, $\hat{n}_j$, of the sky map, and compute the total number of pixels with power exceeding half the maximum in order to measure the beam size over the full $S_2 \times S_2$ space. However, it turns out that regardless of the choice of $\hat{n}_j$, the Poincar\'e point-source response always has $\Omega^{\rm pol}_{\rm p.s.} = 2 \pi\,{\rm sr}$, and so the size of the dirty point-source response can be factorized into its size on the sky and the Poincar\'e sphere.

\begin{figure}
    \centering
    \includegraphics[width=\columnwidth]{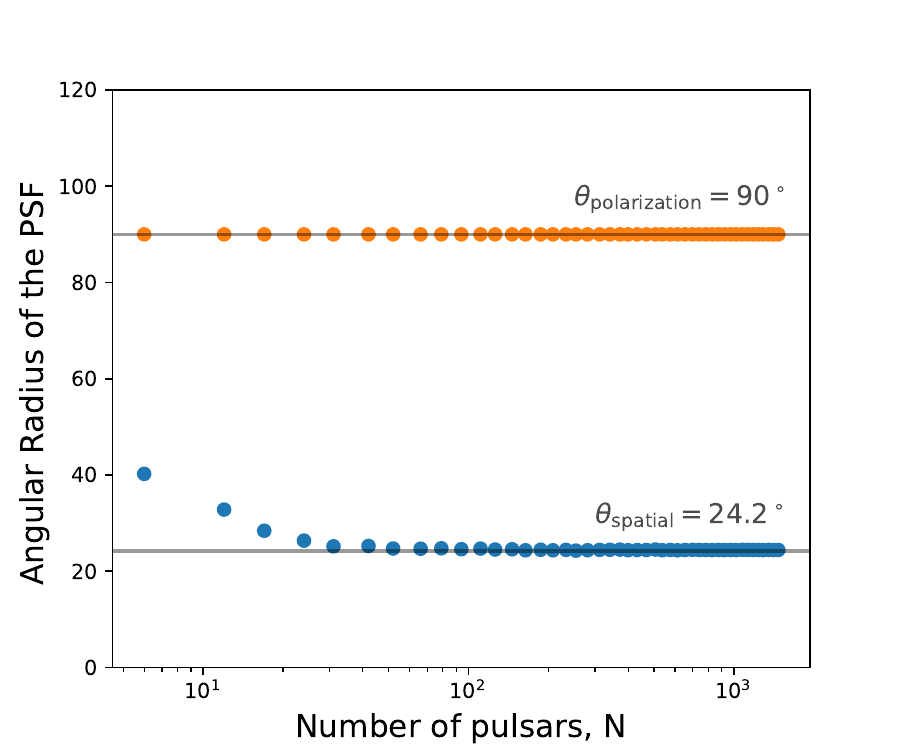}
    \caption{The angular radius of the dirty Stokes I and Poincar\'e maps for a circularly-polarized point source as a function of the number of uniformly distributed pulsars. The size of the matched-filter response of the PTA is characterized by the area on the sky for which the map exceeds half its maximum, which we convert to an angular radius. The limiting values for large $N$ are shown in grey.}
    \label{fig:resolution_Npulsar}
\end{figure}

If $\Omega^{I}_{\rm p.s.} = 0.56\,{\rm sr}$ and $\Omega^{\rm pol}_{\rm p.s.} = 2 \pi\,{\rm sr}$ characterizes the size of the matched-filter point-source response for a dense PTA, at what point is this limiting size achieved? To assess this question we compute $\Omega^{I}_{\rm p.s.}$ and $\Omega^{\rm pol}_{\rm p.s.}$ as a function of the number of pulsars in the PTA, $N$. At low $N$, these numbers depend strongly on the precise distribution of the pulsars on the sky relative to the input gravitational wave source direction (for example, a PTA with pulsars only in the northern hemisphere can better resolve sources emanating from the northern hemisphere than the southern). Thus, to compute the size of the point-source response as a function of the number of pulsars, we construct a PTA by distributing pulsars uniformly on a rectangular grid in $(\theta, \phi \sin\theta) \in [0, \pi) \times [0, 2\pi)$ and removing any values for which $\phi \sin \theta > 2 \pi$. This results in a PTA with pulsars distributed roughly uniformly over the sky. We can increase the number of pulsars incrementally by decreasing the grid spacing. Then, we compute the average value of $\Omega^{I}_{\rm p.s.}$ and $\Omega^{\rm pol}_{\rm p.s.}$ by averaging over their values for random positions of the gravitational wave source, $\hat{n}$, on the sky, and converting the resulting value to an angular radius, $\theta_{\rm spatial}$ and $\theta_{\rm polarization}$. Figure~\ref{fig:resolution_Npulsar} shows the result of this calculation. From this, we can see that the limiting response size is achieved at $N \approx 20$ pulsars.

\begin{figure}
    \centering
    \includegraphics[width=\linewidth]{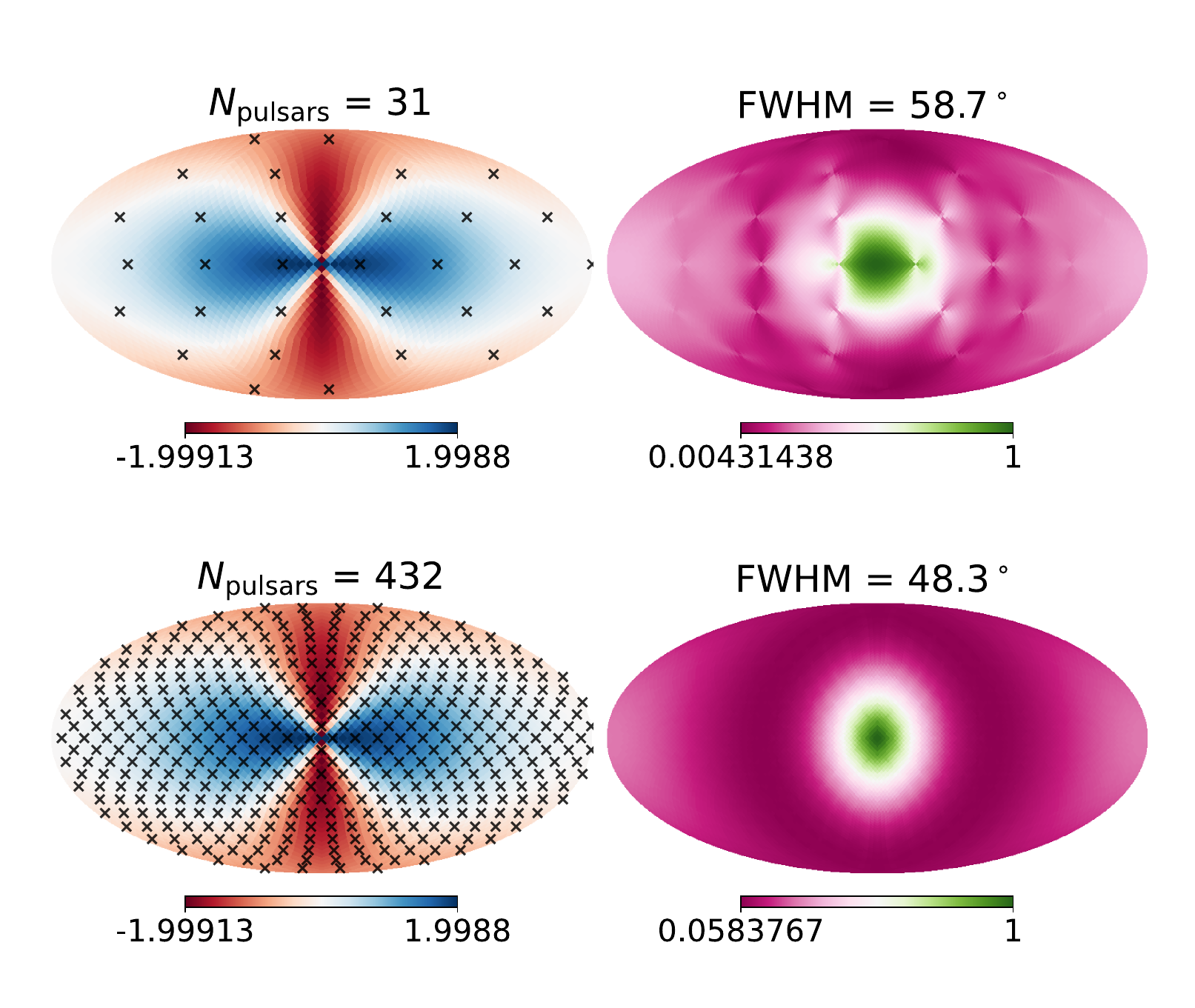}
    \caption{The matched-filter response to a point source of two PTAs with $N=31$ pulsars (top) and $N=432$ pulsars (bottom). The maps on the right shows the right show the resulting Stokes I response in map-space. The maps on the left show the timing residuals induced by the gravitational wave source as a function of pulsar position. The crosses show the actual positions of the pulsars in the simulated PTA. In other words, the maps on the right are produced by sampling the PTA response on the left, at the positions defined by the crosses.}
    \label{fig:resolution_example}
\end{figure}

Figure~\ref{fig:resolution_example} shows a clear example of this phenomenon. The two rows show the Stokes I and Poincar\'e point-source response on the sky for a gravitational wave located at $\hat{n} = \hat{x}$ for two different PTAs with $N=31$ and $N=432$, respectively. The left maps show the continuous response of a hypothetical, infinitely dense PTA induced by the source, and the crosses show the location of the pulsars in the actual PTA we are simulating. In other words, the crosses show the points at which the response function is sampled and the maps on the right show the resulting Stokes I map-response generated from those samples. As can be clearly seen, despite the fact that the $N =432$ PTA more finely samples the response, the resulting point-source response has essentially the same extent as the $N=31$ PTA. Thus, despite increasing the number of pulsars by more than an order of magnitude, the effective independent pixels in the dirty maps remains constant.  

\subsection{Map noise properties and natural maps} 
\label{sec:natural}

We have now computed the matched-filter point-source response of the dirty Stokes I maps and found that it has a limiting spatial radius of $24.2^\circ$. One might take this to be the limiting spatial resolution of a dense PTA; however, characterizing the resolution of the maps is complicated by the fact that noise is correlated across pixels. Consider a noise-only PTA response, $\mathbf{d}^{\rm noise} = [\tilde{\tau}_1, ...,\tilde{\tau}_N]$, where we assume each $\tilde{\tau}_i$ is an independent, complex Gaussian random variable. Then the complex dirty maps, $\mathbf{L}^{\rm noise}, \mathbf{R}^{\rm noise} = (A^{L, R})^\dagger \mathbf{d}^{\rm noise}$ are Gaussian random fields with a non-trivial power spectrum. That is, even though the noise of the timing residuals is uncorrelated across pulsars, multiplying the aperture matrices induces correlations in the noise between pairs of pixels. One can show that the angular power spectrum for the noise is related to the angular power spectrum of a point-source by $C^{\rm noise}_\ell \propto \sqrt{(2 \ell + 1) C_\ell^{|L|}}$, where $C^{|L|}_\ell$ is the power spectrum computed for the point-source response of the dirty $|L|$ map. Note that the power spectrum $C^{|L|}_\ell$ is the same as the power spectrum for the point-source response of the dirty $|R|$ map, $C^{|R|}_\ell$. We introduce what we will call the ``natural" maps, defined by:
\begin{align}
    \hat{L}_j &\equiv \sum_{\ell = 0}^\infty \frac{a^L_{\ell m}}{\sqrt{C^{\rm noise}_\ell}} Y^m_\ell (\hat{n}_j) , \\
    \hat{R}_j &\equiv \sum_{\ell = 0}^\infty \frac{a^R_{\ell m}}{\sqrt{C^{\rm noise}_\ell}} Y^m_\ell (\hat{n}_j).
    \label{eq:natmap}
\end{align}
where $a^{L, R}_{\ell m}$ is the spherical harmonic decomposition of the dirty $L$ and $R$ maps. In other words, the natural maps are the dirty maps weighted by the inverse square-root of the noise power spectrum. This procedure, when applied to the pure noise maps, $\mathbf{L}^{\rm noise}$, $\mathbf{R}^{\rm noise}$, produces natural maps with a flat power spectrum, i.e. Gaussian and uncorrelated noise. The point-source response computed for the natural maps is the true beam of the experiment, i.e. the point-spread function (PSF) in the space with uncorrelated, Gaussian noise. 

\begin{figure}
    \centering
    \includegraphics[width=\linewidth]{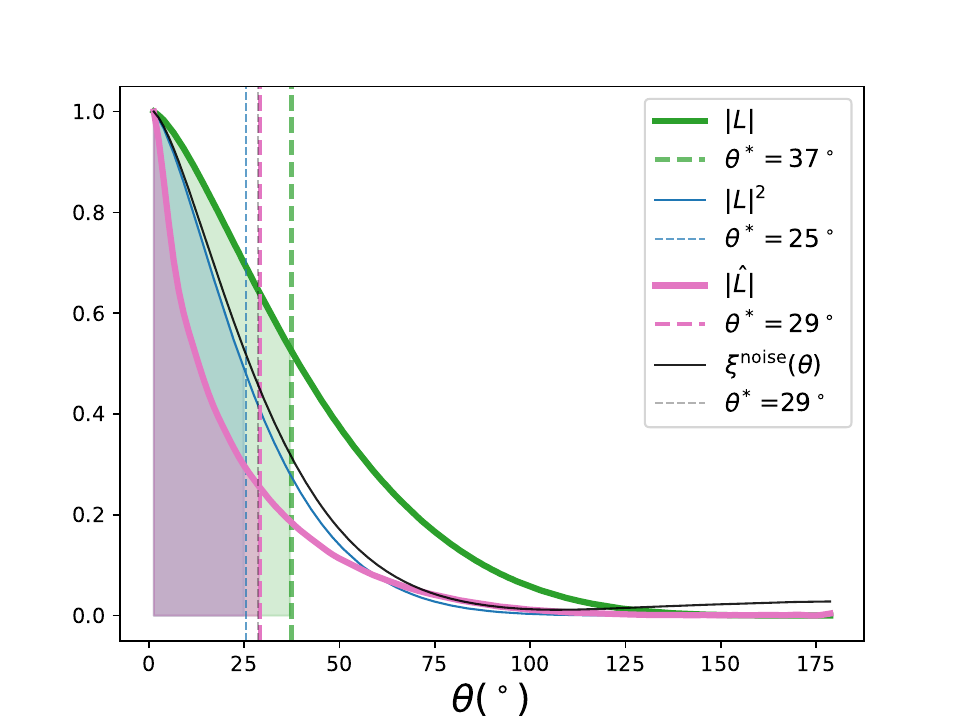}
    \caption{Radial profiles of $|L_j|$, $|L_j|^2$, and $|\hat{L}_j|$ (defined by Eq.~\ref{eq:natmap}) for a purely left-circularly-polarized point source located at the North pole.  Also shown is the angular two-point correlation function for a pure noise map $\mathbf{L}^{\rm noise}$. Widths of the radial profiles as defined by Eq.~\ref{eq:fwidth} are shown as dashed vertical lines. The profile of the natural map, $\hat{L}$, can be regarded as the beam of the experiment, as it is the PSF in the space where the pixel-noise is Gaussian and uncorrelated.}
    \label{fig:radial_profile}
\end{figure}

Figure~\ref{fig:radial_profile} shows a comparison of the radial profiles of the natural $|\hat{L}_j|$ map (the beam), the dirty $|L_j|$ map (the filtered beam), and the dirty $|L_j|^2$ map for a purely left-circularly-polarized source at the pole as a function of $\theta$. We also show the angular two-point correlation function for the dirty map noise, $\xi^{\rm noise} = \sum_\ell^\infty \frac{2 \ell + 1}{4 \pi} C^{\rm noise}_\ell P_\ell (\cos\theta)$. All of the profiles are normalized so that their maximum is unity. Note that $|L_j|^2$ profile is simply the radial profile of the intensity response computed in the previous section and shown in Fig.~\ref{fig:resolution_example}.

In order to characterize the resolution of the PTA, we would like to compare the widths of these radial profiles. Since the natural map profile has a significant cusp compared to the other profiles, instead of defining the width by the half-maximum point, we define the angular size of a radial profile $f(\theta)$ by the angle $\theta^*$ satisfying
\begin{equation}
    \int_0^{\theta^*} f(\theta) \sin \theta d\theta = 0.38 \int_0^\pi f(\theta) \sin\theta d\theta.
    \label{eq:fwidth}
\end{equation}
For a Gaussian profile, this is equivalent to the half-width, $\sigma$. Defined in this way, we see that the effective radius of the intensity profile, $|L|^2$, is $\theta^* = 25^\circ$, roughly consistent with what was found for the dirty intensity responses in the previous section. Despite the sharper profile, the natural map has a slightly larger effective radius of $\theta^* = 29^\circ$, which is the same as the effective radius of the noise two-point correlation function, $\xi^{\rm noise}(\theta)$. We take the width $\theta_{\rm res} = 2 \theta^* = 58^\circ$ to be the effective spatial resolution of a dense PTA. We arrived at this number in essentially two ways: it is the effective angular width of the two-point correlation function of the purely noise dirty maps, $\mathbf{L}^{\rm noise}$, $\mathbf{R}^{\rm noise}$, or, alternatively, it is the width of the PSF in the whitened, ``natural" map, $\hat{L}$. 

We can convert this spatial resolution into the number of distinct
positions of point-sources a dense PTA can effectively distinguish:
$N_{\rm spatial} = \left. 4 \pi \right. / \left. 2 \pi \int_0^{\theta^*} \sin\theta
  d\theta \right. \approx 16$. Alternatively, this can be taken as the
effective number of $\textit{independent}$ pixels in both of the
complex $L$ and $R$ maps. A similar analysis can be carried out to
find the effective number of independent pixels in the Poincar\'e map,
resulting in a number of distinguishable polarization states of
$N_{\rm pol} \approx 2$. The final result is that a dense PTA can
distinguish only $N_{\rm res} = N_{\rm spatial} \times N_{\rm pol} =
32$ polarized point sources. This is, of course, a rough estimate that
is signal-to-noise dependent; with infinite signal to noise, any
number of sources can always be resolved regardless of the beam
width. The scaling of the resolution with signal-to-noise depends on
the precise shape of the beam, with sharper beams yielding finer
resolutions.   

\section{\label{sec:HD}Relationship to Hellings-Downs}

We would now like to compare the information content of the polarized maps with the standard Hellings-Downs analysis. To start, we define the covariance matrix:
\begin{align}
    C &\equiv A^L (A^L)^\dagger + A^R (A^R)^\dagger, \\
    C_{i\ell} &= 2 \sum_{j=1}^M \zeta_{ij}\overline{\zeta}_{\ell j},
\end{align}
where, for simplicity, we've defined $\zeta_{ij}\equiv\zeta^{L,R}(\hat{r}_i;\hat{n}_j)$, suppressing the polarization labels, $L, R$ (since $\zeta^L$ and $\zeta^R$ are complex conjugates of each other, $C_{i\ell}$ does not depend on the choice of polarization). Averaging these correlations for a fixed angular separation between pulsar pairs yields
\begin{align}
    &\langle C_{i\ell}\rangle \big|_{\hat{r}_i \cdot \hat{r}_\ell = \mu} = 8 M \Gamma(\mu), \\
    &\Gamma(\mu) = \frac{1}{3} \left\{ 1 + \frac{3}{2} (1 - \mu) \left(\log(\frac{1 - \mu}{2}) - \frac{1}{6} \right)\right\},
\end{align}
where $\Gamma(\mu)$ is the well-known form of the Hellings-Downs curve (also known as the overlap reduction function for an isotropic background) and $\mu = \cos\theta$, where $\theta$ is the angular separation between pulsar pairs. A similar argument for the correspondence between the covariance matrix and the overlap reduction function can be found in \citep{2020PhRvD.102h4039T}. This result is unsurprising as the elements of $C$ are correlations of pulsar pairs, and the averages of those correlations for fixed angular separation is precisely what the Hellings-Downs curve represents.\footnote{Note that a second covariance matrix can be defined: the $N \times N$ matrix $C_{\rm pixel} \equiv  (A^L)^\dagger A^L  + (A^R)^\dagger A^R $.  Each column of this matrix is a dirty map for a single GW with different positions and polarizations.  The trace of $C_{\rm pixel}$ is the sum of $|\zeta(\hat{r}_i;\hat{n}_j,\hat{n}_p)|^2$ over all values of $(i,j,p)$ and is equivalent to the trace of $C$.  Averaging the pixel-pixel correlations for fixed angular separation, one obtains the noise two-point correlation function, $\xi^{\rm noise}(\theta)$ (shown in in Fig.~\ref{fig:radial_profile}), which is distinct from the Hellings-Downs curve.  This is not unexpected since $A$, $A^{\dagger} A \neq A A ^{\dagger}$. In particular, convolutions do not generally commute on a sphere.  However, in the flat-sky approximation, $A$ and $A^\dagger$ \textit{do} commute, and we would recover the Hellings-Downs curve in the pixel-pixel basis.}

In light of this, we can rewrite the total power contained in the dirty intensity map:
\begin{equation}
    \sum_{j=1}^M I_j = {\bf L}^\dagger {\bf L} + {\bf R}^\dagger{\bf R} = {\bf d}^\dagger (\Sigma^{-1})^T C \Sigma^{-1}{\bf d} = \sum_{i,\ell = 1}^N \frac{d_i \overline{d}_\ell C_{i \ell}}{\sigma_i \sigma_l},
\end{equation}
where the last equality is obtained under the assumption that $\Sigma_{ij} = \sigma_i \delta_{ij}$. In other words, the sum over intensity in map-space can be related to a sum over pulsar-pairs, weighted by their expected isotropic correlation, $C_{i\ell}$. Note that this sum is precisely the so-called ``optimal statistic" frequently employed to assess the significance of the observed GWB \citep{2009PhRvD..79h4030A, 2015PhRvD..91d4048C, NANOGravdetection}. Thus, summing the power in the map is identical to the standard Hellings-Downs statistical analysis for an isotropic background.

Noting the equivalence of summing the total map power to the ``optimal statistic", the advantage of map-making approaches becomes clear. The summed intensity, or equivalently the map monopole, manifestly contains the same information as the Hellings-Downs weighted timing-residual cross-correlations. Moreover, the maps encode additional anisotropy and polarization information. While the ``optimal statistic" is specifically optimized for a true stochastic and statistically isotropic background, map-making is a more general technique that will outperform the optimal statistic when the GWB is dominated by a small number of bright point sources. The reason for this is straightforward: for a map dominated by a few bright pixels, the signal-to-noise ratio of the brightest pixel will always be greater than the signal-to-noise of the total intensity, where the signal has been diluted by summing over noisy pixels. 

For a more quantitative comparison, consider the extreme case of a gravitational wave ``background" comprising a single, left-circularly polarized source. Following our map-making procedure, we can produce two complex maps: $L_j$, and $R_j$. As discussed in Section~\ref{sec:resolution}, these maps can distinguish at most $N_{\rm res} = 32$ distinct polarized sources. Alternatively, the $L_j$ and $R_j$ maps have 32 independent pixels between them. The number of independent pixels is crucial for determining the look-elsewhere factor for assessing the detection significance. Now, consider a noisy  measurement, such that the observed response at each pulsar, $d_i$, is a complex, Gaussian random number. The result is that, in the absence of any signal, for each pixel we have: ${\rm Re}[L_j],{\rm Im}[L_j], {\rm Re}[R_j], {\rm Re}[R_j] \sim \mathcal{N}(0,\sigma^2)$. It immediately follows that, in the absence of any signal, the pixel values of the left- and right-intensity maps are $\chi^2$-distributed: $|L_j / \sigma|^2, |R_j / \sigma|^2 \sim \chi^2(2)$. Likewise, the value of the sum over the intensity map is $\sum_{j=1}^{N_{\rm res}/2} I_j / \sigma^2 \sim \chi^2(2 N_{\rm res})$, where the sum is computed over the total number of independent spatial pixels, $N_{\rm res} / 2 = 16$.

Now, we can compute the significance of a signal from a single, left-polarized source, such that the pixel in the direction of the source attains an amplitude of $|L_{\rm max}|$ and the other pixels are distributed according to their null-distributions. The signal-to-noise in the maximum pixel is simply $|L_{\rm max}| / \sigma$. However, since we want to assess the probability that \textit{any} of the independent resolution elements could have attained that value, we must account for the look-elsewhere effect. Thus, we compute the $p$-value via $p = P_{\chi^2(2)} (X > |L_{\rm max} / \sigma|^2) \times N_{\rm res}$, where we multiply the probability that a given $\chi^2(2)$-distributed element would exceed the observed signal value by the number of independent elements. We can then take this $p$-value and convert it to a standard Gaussian signal-to-noise via
\begin{equation}
    \mathrm{(S/N)}_{\rm pixel} = - \sqrt{2} {\rm erf}^{-1}( 2 N_{\rm res} e^{-\frac{1}{2}|L_{\rm max} / \sigma|^2} - 1),
\end{equation}
where we have taken advantage of the fact that the cumulative distribution function for a $\chi^2(2)$ distribution is $1 - e^{-x/2}$.

To compare with the standard ``optimal statistic", we compute the probability that the sum of total intensity for a noise-only map would exceed the actually observed value. Since the actual sum includes noise pixels, the observed value depends not only on the signal amplitude, but also on the actual noise realization. To estimate the significance, we simply take the sum over the non-signal pixels to be the mean of the $\chi^2(N_{\rm res})$ distribution. Thus, the $p$-value of the summed intensity is $p = P_{\chi^2(2N_{\rm res})}(X > |L_{\rm max}/\sigma|^2 + {\rm E}[\chi^2(2 N_{\rm res})])$. Since a $\chi^2$ distribution with $2 N_{\rm res} = 64$ degrees of freedom is approximately a Gaussian, $\mathcal{N}(2 N_{\rm res}, 4 N_{\rm res})$, the Gaussian signal-to-noise of the summed intensities is simply
\begin{equation}
    \mathrm{(S/N)}_{\rm H.D.} \approx |L_{\rm max} / \sigma|^2 / \sqrt{4 N_{\rm p.s}}.
\end{equation}

Putting some actual numbers in now, let $|L_{\rm max}|=1$ and $\sigma=0.16$. By summing the intensities, one attains a detection significance of $3 \sigma$. However, by computing the significance of the maximum pixel, one attains a detection significance of $5.2 \sigma$. Thus, a marginal detection in total intensity can be a significant detection in map-space, in the regime where a small number of sources dominate the signal. In doing this calculation, we have assumed that there are exactly $N_{\rm res} = 32$ pixels across the sky and polarization, with identically and independently distributed noise properties. Of course, in reality, in order for the pixels to be truly identical, one must be in the limit of an infinite number of pulsars, distributed uniformly on the sky, so that the PTA sensitivity is not direction dependent. Even in that limit, the pixels are never truly independent, as the PSF has support across the entire sky. Nevertheless, this simple calculation gives a simple benchmark for comparison. A more detailed comparison, taking these subtleties into account, will be the subject of future work.

\section{\label{sec:discussion}Discussion}
 
The main argument of this paper is twofold: 1. there is a fundamental limit to the number of polarized sources that a PTA (with unknown pulsar distances) can resolve, and 2. if the GWB is dominated by resolvable sources, then polarized map-making techniques have greater statistical power than the standard, unpolarized, and isotropic Hellings-Downs analysis. A natural question is whether or not the observed nanohertz GWB is, indeed, likely to comprise a distribution of resolvable sources, the brightest of which might be individually detected.

Ref.~\cite{Sato-PolitoZaldarriagaQuataert2023} have argued that the supermassive black hole (SMBH) population density inferred from the actually amplitude of the GWB is an order of magnitude higher than the predicted value based on our current understanding of SMBH populations. One simple resolution to this tension is that we happen to be nearby a particularly bright source or sources. If so, the observed amplitude would be much larger than the average expected amplitude, and the nanohertz gravitational wave sky would be dominated by a few, resolvable sources. Simulations show that this is not a highly unlikely scenario, at least for some frequency channels. Indeed, Fig. 1 of Ref.~\cite{Kelley2018} suggests that it is not uncommon in simulations of the astrophysical background for the brightest source in a given frequency channel to be an order of magnitude brighter in intensity than the rest of the sources combined. Of course, such a scenario would have a very different spectral distribution of the observed amplitude compared to an isotropic GWB. The current constraints of the spectral distribution are consistent with the expected power law for a purely isotropic background \citep{NANOGravdetection}. However, we argue that the current observations are not particularly constraining; the NANOGrav analysis, essentially, only has four frequency bins with any constraining power, spanning about a factor of 5 in frequency. Even modest initial eccentricities for SMBH mergers can produce large spreads in the spectral energy distribution of a single source \citep{1963PhRv..131..435P, 2007PThPh.117..241E}. 

Nevertheless, searches for individual, bright sources in the PTA data have been inconclusive. Ref.~\cite{NANOGrav_CW} perform a search for individual sources in the NANOGrav 15-year dataset and do not find significant evidence for the presence of a single, bright source. However, it should be noted that this analysis is performed by comparing Bayes' factors of four models: 1. no gravitational waves, 2. a single source, 3. pure, isotropic Hellings-Downs correlations, and 4. a single source plus pure Hellings-Downs correlations. They find that the data is consistent with pure, isotropic Hellings-Downs correlations alone. However, since a single source already produces Hellings-Downs-like correlations on its own, the consistency of the data with isotropic Hellings-Downs correlations should not be taken as ruling out the possibility that the observed background comprises a handful of resolvable sources. If, indeed, the background is the result of a small number of sources, this would result in a highly anisotropic background. Ref.~\cite{Nanograv2023Anisotropy} perform a search for anisotropy in the background, but again, do not find a significant evidence of anisotropy. Note, however, this analysis searches for evidence of anisotropy in the angular power spectrum of the gravitational wave power on the sky. This is certainly not optimal for the kind of anisotropy that would be induced by a small number of resolved point sources. Ref.~\cite{Nanograv2023Anisotropy} also produce intensity maps of the kind we have described here and search for individual pixels that rise above some signal-to-noise threshold (see Fig.~9 of \cite{Nanograv2023Anisotropy}). However, because they assume the spatial pixels are independent, it is difficult to interpret the true significance of these maps without a more detailed analysis of the noise properties of the pixels. 

Thus, while the data, at present, \textit{favours} an isotropic GWB, the possibility that the observed background also contains a small number of bright sources has not been ruled out. Moreover, assuming a purely isotropic background results in an observed amplitude that is in tension with the predicted amplitude. However, searches for resolvable, coherent, and polarized sources have not been optimized. Simple quadratic estimates of the polarized power, such as we have presented here, can be leveraged to perform those searches in a way that does not rely on complex model comparison. This approach allows for a simultaneous fitting of a bright, resolved source with a random residual background\citep{2003ApJ...591..575M}. Further work is needed to understand the detailed noise properties of these polarized maps for realistic PTAs.

\section{Conclusion}
In this paper, we have presented a construction for polarized maps of the nanohertz gravitational wave sky. By computing the point-spread function of these maps, we have demonstrated a fundamental limit to the resolving power of PTAs without well-characterized pulsar distances. Specifically, without pulsar distances, even an infinitely dense PTA can, at most, distinguish $N_{\rm res} = 16 \times 2$ distinct, polarized point-sources. Moreover, we have shown that the monopole of the intensity map is equivalent to the ``optimal statistic" commonly employed to detect the presence of an isotropic gravitational wave background. Thus, since polarized maps contain additional information on anisotropy and polarization, we argue that it is always advantageous to transform PTA observables into polarized map-space. In particular, given that astrophysical origins of the background are expected to comprise an ensemble of individual, nearly monochromatic, and polarized point-sources, coherent map-based techniques may have significantly greater statistical power than the so-called ``optimal statistic". While more work is needed to characterize the noise properties of PTA maps in order to employ them for robust gravitational wave detection, this work can be motivated by the evident advantages of map-based approaches. In addition to the statistical advantages, robust sky-maps may provide more accessible means of interacting with data products in public data releases, enabling greater engagement from the broader community on high-level analyses. The historical precedence of CMB studies may be taken as further evidence of the utility of map-based approaches over direct time-stream analyses. 

\begin{acknowledgments}
U.P. is supported by the Natural Sciences and Engineering Research Council of Canada (NSERC) [funding reference number RGPIN-2019-06770, ALLRP 586559-23, RGPIN-2025-06396] , Canadian Institute for Advanced Research (CIFAR), Ontario Research Fund (ORF-RE Fund), and AMD AI Quantum Astro.
\end{acknowledgments}

\appendix

\nocite{*}

\bibliography{biblio}

\end{document}